# Interpretation of Past Kingdom's Poems to Reconstruct the Physical Phenomena in the Past:

# Case of Great Tambora Eruption 1815

Mikrajuddin Abdullah

Department of Physics, Bandung Institute of Technology,

Jl. Ganesa 10 Bandung 40132, Indonesia

and

Bandung Innovation Center

Jl. Sembrani No 19 Bandung, Indonesia

Email: mikrajuddin@gmail.com

**Abstract**

In this paper I reconstruct the distribution of ash released from great Tambora eruption in 1815. The reconstruction was developed by analyzing the meaning of *Poem of Bima Kingdom* written in 1830. I also compared the effect of Tambora eruption and the Kelud eruption February 13, 2014 to estimate the most logical thickness of ash that has fallen in Bima district. I get a more logical value for ash thickness in Bima district of around 60 cm, not 10 cm as widely reported. I also propose the change of ahs distribution. Until presently, it was accaepted that the ash thickness distributiom satisfied asymmetry cone distribution. However, I propose the ash distributed according to asymmetry Gaussian function. Random simulation and fitting of the thickness data extracted from isopach map showed that the asymmetry Gaussian distribution is more acceptable. I obtained the total volume of ash released in Tambora eruption was around 180 km$^3$, greater of about 20% than the present accepted value of 150 km$^3$.

## I. INTRODUCTION

What the scientists do to explain phenomena in the past during which science was unde developed? What the scientists do to reconstruct phenomena in the past at locations where scientific activities were under established? A most common approach is by excavating the location at which the events have happened and analyzing the collected results using modern science methods. Radioactive dating techniques, the method introduced by Willard Libby in 1940s [1,2], have been often carried out to investigate fossils or other burried objects. Another possible approach is to study documents written by local residents that have lived around location where the event has happened, in particular documents written at time when the event happened or not long after the event happened. The document can be poetic or other styles.

One catastropic phenomenon in the past having little direct scientific observations is the great Tambora eruption 1815. This eruption is believed to cause the year without summer in northern hemisphere. The Tambora eruption in Sumbawa island, Indonesia ($8.2^oS$, $118^oE$) at 11 April 1815 is the largest volcanic eruptions recorded in human history [3]. Although there were another greater eruptions before, such as Samalas (Rinjani) eruption in Lombok island, Indonesia in 1257 [4], Yellowstone eruption in Wyoming, USA 640,000 years ago, Taupo eruption in New Zealand 26,500 years ago, and Toba eruption in Sumatera, Indonesia 74,000 years ago [5], the eruption of Tambora becomes interesting because it happened in the modern era and several phenomena accompanying the eruption have been perceived and recorded by modern civilization.

In 2015, the 200th anniversary of Tambora eruption was celebrated in Dompu district, West Nusa Tenggara, Indonesia. The western part of mount Tambora is located in Dompu district while the eastern part is located in Bima district, but the kingdoms of Tambora and Pekat that have been were disappeared by the eruption are located in Dompu district. In 2014, I was asked by some officers of Dompu district to make a popular scientific literature regarding the Tamborain eruption. I read a lot of literature related to the eruption, and most of them are publications in journals of geology or volcanology. During the work I found some debatable conclusions in published works. The first is the prediction volcano ash thickness in Bima district. In my opinion, the thickness of ash in Bima district of about 10 cm was likely too thin. The second thing is the shape of ash distribution to follow asymmetry cone which has been used as a model for calculating the volume volcano ash waslikely not so accurate. My first opinion is the thickness of ash that has fallen in Bima district must be much larger than 10 cm by considering the fact that the distance of Bima to Tambora is only 90 km, and the Tambora eruption was the largest eruption in recorded human history. Moreover, the Tambora eruption processed fo nearly within one week, even though the largest eruption has occurred at 11 April 1815. My second opinion is the ash distribution not in cone shape, but likely a normal distribution. It is very known in scientific reasearch that nearly all natural process would tend to follow the normal or log normal distribution.

The purpose of this paper is to reconstruct the distribution of Tambora eruption ash and the thickness of the ash that fell in Bima district. From these two results, I repredict the volume of ash from the Tambora eruption.The message that I want to prompt here is how we interpret literature documents in the past and compare with similar events that occurred at present to estimate the quantity of Tambora eruption ash that has fallen in Bima district. The past literature document I have used is the Poem of Bima Kingdom. This poem describes the events that have occurred in the Bima Sultanate at period of 1815-1829. There are four events described in the poem: the death of sultan, appointment of his successor, pirate attack, and Tambora eruption. The poem was written by Lukman Hakim at around 1830 [5].

**II. THE THICKNESS OF ASH FELT IN BIMA MIGH BE TOO THIN?**

Mount Tambora occupies most of the Sanggar peninsula (**Figure 1**).There were little scientific studies short after the eruption because of the difficulty to reach the location which was under the Dutch colonial (now Indonesia). Moreover, the mount Tambora was very far from Batavia (Jakarta), the center of Dutch colonial administration. Some documents regarding the Tambora condition aftereruption were hand written documentsof some England people who lived in the Dutch East Indies, notes of the ship's captain, and soldier [6]. Before eruption, the heigh of the montain was about 4,300 m above sea level, and now the mountain heigh is only 2,850 m.

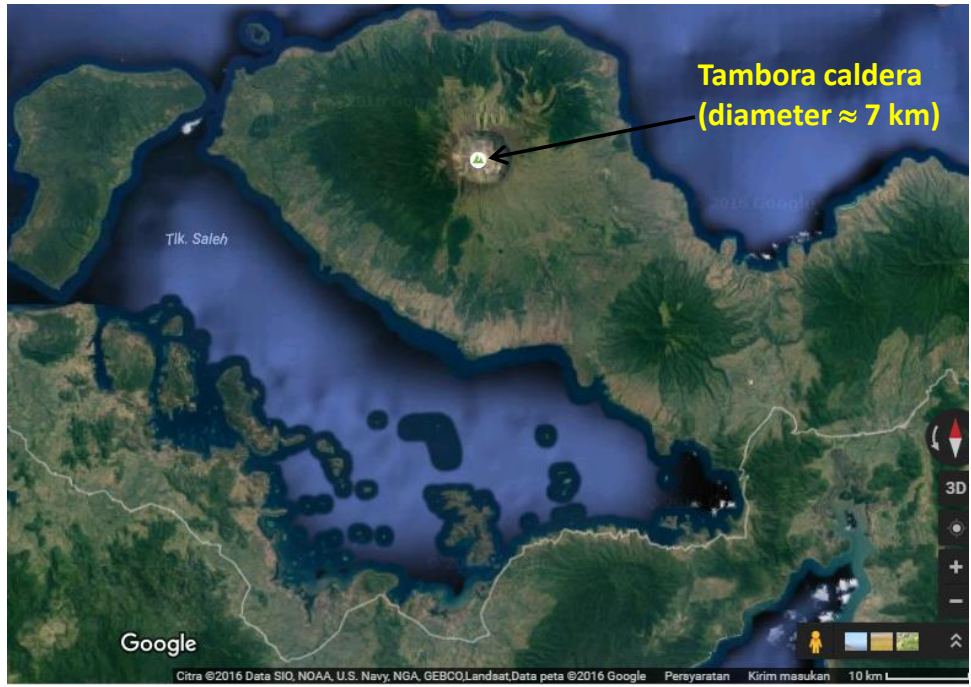

**Figure 1** Location Mount Tambora in Sanggar peninsula, Sumbawa island, Indonesia. Before eruption, the mountain height was about 4,300 m above sea level. Now, the height of the mountain is only 2,850 m above sea level. The eruption produced a caldera of diameter about 7 km [7].

One phenomenon that has been widely studied regarding the Tambora eruption is distribution of volcanic ash. The amount of ash released during eruption can be used to predict the total volume of material produced from the eruption. Tambora eruption caused the ash to spread to most parts of Indonesia. Most ash has fallen in the area of 4.5 x $10^5$ km$^2$ [8]. **Tabel 1** is the recorded ash thickness in several locations in Indonesia after eruption.

**Tabel 1**. The predicted ash thickness at several locations in Indonesia after Tambora eruption [8].

| Location | Distance from caldera (km), direction | Ash started to fall | Final thichness (cm)* |
|---|---|---|---|
| Tambora | < 20 | 10 April, 7 PM | 90 |
| Sanggar | 30, east | 10 April, 10 PM | thick |
| Bima | 90, east | 11 April, 7 AM | 10 |
| Makassar | 380, north | 11 April morning | 4 |
| Sumbawa Besar | 60, west | - | thick |
| Lombok | 200, west | - | 60 |
| Bali | 300, west | morning | thick |
| Banyuwangi | 400, west | 13 PM | 22 |
| Panarukan | 450, west | - | 5 |
| Sumenep | 470, west | 4 PM | 5 |
| Besuki | 500, west | 4 PM | 5 |
| Probolinggo | 530, west | - | 5 |
| Surabaya | 590, west | 10 night | little |
| Gresik | 600, west | night | little |
| Rembang | 740, west | - | little |
| Surakarta | 790, west | - | little |
| Jogjakarta | 830, west | - | little |
| Semarang | 840, west | - | little |
| Tegal | 990, west | - | little |
| Batavia | 1,260, west | - | little |
| Banten | 1,330, west | - | little |

**Figure 2** is the isopach map of Tambora ash distribution. It must be noted that the data were generally reported by local residents, officers in Dutch collonial, and not based on direct measurement by scientists at that time [8]. Scientists have predicted the volume of materials produced by Tambora eruption, and the figures varied very wide such as 300 km$^3$ [9], 1,000 km$^3$ [10], 150 km$^3$ [11], and 100 km$^3$ [12]. If the calculation is based on the volume of Tambora top that diasappeared due to eruption, the estimated volume is only 30 km$^3$ [13,14].

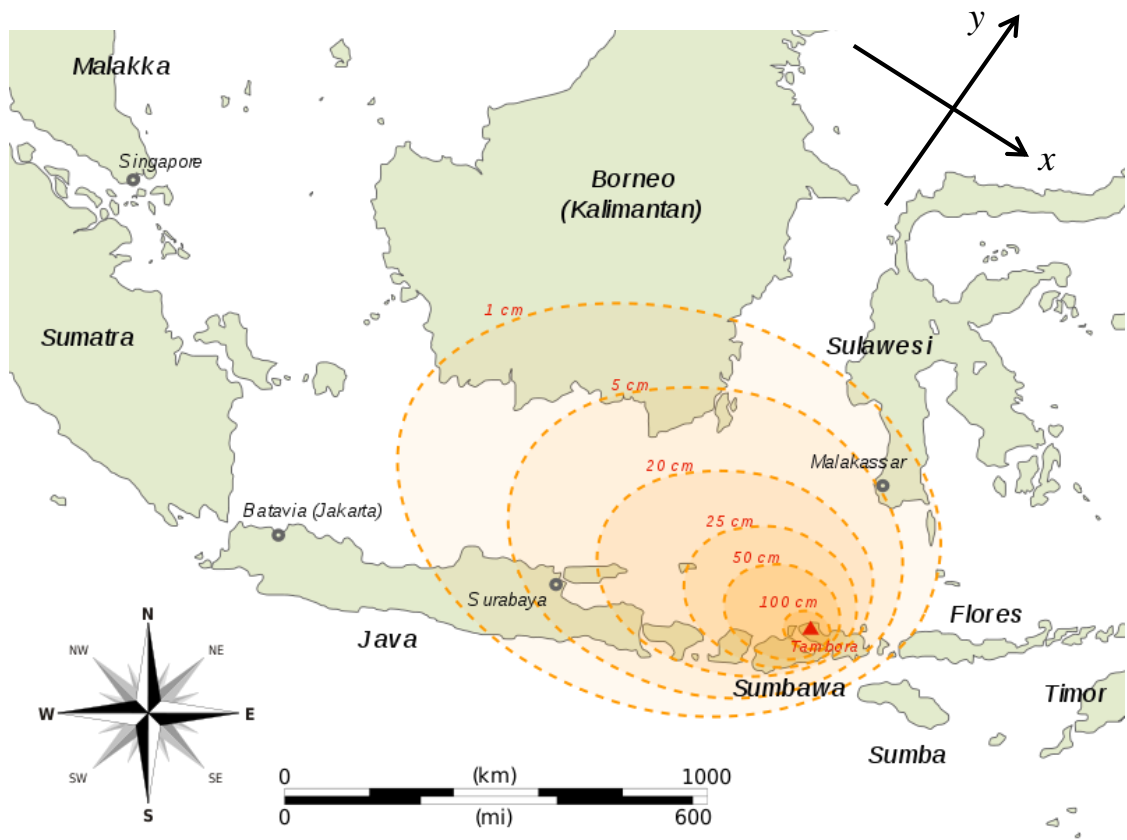

**Figure 2** Isopach map of Tambora ash [15]. The isopach curves are not circles, but likely ellipses stretched toward west-northern direction. For calculation purpose we choose x axis directs parallel to longer axis of the ellipses (see inset).

It is clear from the isopach map or data in **Tabel 1** that the volcano ash felt in Bima district is too thin, around 10 cm, although the distance of Bima district to Tambora caldera is only around 90 km. Asymmetry distribution was caused by easterly monsoonwinds [16] flown to west-northern after eruption such that the ash has fallen at locations far to west-northern direction up to Borneo (now Kalimantan) and little amount felt to southeastern. Based on Stother analysis, during Tambora eruption the wind velocity was as following. There was a steady velocity of 3 m/s toward west-northern and a random velocity of 4 m/s [8]. However, there is still a question: was not the ash felt in Bima district too thin? If the ash thickness in Bima district is thicker that the value accepted at present, the volume of material produced by Tambora eruption becomes larger than the value accepted at present.

In several publications, the authors rarely referred the documents of Bima Kingdom. Bima Kingdom was the influenced sultanate in Sumbawa island during that period, and also has annectation region in Sumba island. The most frequently cited document was the Zollinger publication in 1855, after he visited Tambora in 1847. In the documents of Bima Kingdom, there was a *Poem of Bima Kingdom* expalaining the Tambora eruption. If we carefully analize the message of the poem, it becomes clear that effect of Tambora eruption in Bima district was very catastrophic, so the amount of ash felt in Bima district should be thicker than the accepted figures at present.

To imagine the condition of Bima district following the Tambora eruption we can carefully read a note by Francis after visited Bima in 1831 [17] as rewritten by Chambert-Loir [18]. Francis wrote: "Tambora eruption resulted catastrophic effect: the land was covered by ash as thick as 2 feets during five days, a lot of houses destroyed, all crops tottaly destroyed. The land could not be vegetated for five years. Great famine occurred. The rice was imported from Java with a price of 8 gulded for each yoke. The people were so miserable, all family relationships severed; there were husbands sold their wives, there were mother sold their children, bartered with little food; poor people died in the streets." It is clear from the note that the thickness of ash in Bima distric was 2 feets or around 66 cm.

In addition, written in *Poem of Bima Kingdom*, there are several strophes expressing the catastrophic descturction. **Table 2** lists some strophe numbers and line in the strophe explaining condition of Bima district after Tambora eruption.

**Table 2** Several strophes in the *Poem of Bima Kingdom* expressing the condition of Bima district after Tambora eruption [18]

| Strophe number | Line in the strophe |
|---|---|
| 12 | Bima land scorched all its rice |
|  | everyone in hunger |
| 13 | hunger is extraordinary |
|  | it seems the country is not excited |
|  | The world was likely afterdoomsday |
| 16 | thousands of people died |
| 18 | ash fell two days and three nights |
| 19 | Ash fell so much and pile up |
| 23 | bright day turned dark |
| 24 | ash fell as poured |
| 33 | can not distinguish day and night |
|  | clear weather turned darkens |
| 36 | when the ash fell, the day entered darkness |
|  | lighting lamps at noon |
|  | the objects around were not visible |
|  | walking around like a blind |
| 41 | people walk in groping |
|  | Like fishes that have been poisoned |

Futhermore, Heinrich Zollnger, a Swiss botanist who arrived at Java in 1841 and expedited to Tambora in 1847 wrote a short mogograph in 1855 explained the number of death due to eruption, after eruption, and people emigration as summarized in **Table 3** [10].

**Tabel 3** Number of death and emigration in several locations in Sumbara island due to Tambora eruption [10]

| Location | Killed in the eruption | Died of hunger, or illness | Emigrated |
|---|---|---|---|
| Pekat | 2000(**) | | |
| Tambora | 6000(**) | | |
| Sanggar | 1100 | 825 | 725 |
| Dompu | 1000 | 4000 | 3000 |
| Sumbawa | | 18000 | 18000 |
| Bima | | 15000 | 15000 |

(**) All died during eruption

It is apparent from Table 3 that residents of Tabora and Pekat were killed in the eruption and nono was killed or emigrated after eruption. It indicated the the Tambora and Pekat Kingdoms disappeared after eruption. This condition was also mentioned in Poem of Bima Kingdom at strophe31: become extinct Pekat and Tambora, strophe 32: Pekat andTambora did not survive, andstrophet 34: land and the body melted [18]. It is also apparent from Table 3 that the death toll and emigration in Bima and Sumbawa were nearly the same. It might informed that the effect of eruption felt in Bima and Sumbawa were nearly the same, to mean that the quantity of volcano materials, including the volcano ash, fallen in Bima and Sumbawa were nearly the same. Zollinger also estimated that at least 10,000 also died of starvation and illness on the neighbouring island of Lombok, smaller than the number in Bima district, event Lombok island is more populous than Sumbawa and Bima since long time ago. It mingh informed that the effect of eruption felt in Lombok was slightly less catastrophic than in Bima or Sumbawa. Therefore, the thickness of ash in Bima shuld be thicker that in Lombok. But based on isopach map or Tamble 3 it is clearly seen that ash in Lombok is thicker than in Bima. I guess the thickness of ash in Bima should be thicker than in Lombok, contrary to the data in Table 3 or isopact map in Figure 2.

Other notes explaining catastrophic destruction in Bima after Tambora eruption are as following. Rice started to grow in Bima at 1820, though the recovery was still far from complete. A botanist C. G. C. Reinwardt that visited Bima in 1821experienced great difficulty in getting supplies for his ship because of the dearth of foodstuffs; this contrasted with the situation before 1815, when Bima had been an excellent victualling port. According to Reinwardt, the town was looking derelict and there were hardly any horses— formerly the principal commodity—to be got [19]. The Bima sultanate came to recovery in 1851 [20,21], i.e. 36 years after eruption, indicating a very catastrophic destruction in Bima district.

Based on Bima conditions as mentioned above, it is debatable to accept that the thickness of ash in Bima district was only 10 cm. Moreover the time period of ash falling was very long, ie 2 days and 3 night, or equivalent to around 60 hours.

### III. COMPARISON WITH THE THICKNESS OF ASH FROM KELUD ERUPTION

To support the hypothesis that the thickness of ash in Bima district after Tambora eruption was larger than the presently accepted value, let us analyze the effect of Kelud eruption ($7.9^o$ S, $112.3^o$ E) in East Java, Indonesia that has happed in 2014. The mount Kelud is located at the same ring with mount Tambora. The distance of Kelud and Tambora is around 630 km (data from Google eath). The Kelud eruption happened on February 13, 2014at 22.50 West Indonesia time.Ash from the eruption started to fall in Yogyakarta city, distance of around 215 km from mount Kelud on Feburray 14 at 03.00 West Indonesia time. The thickness of ash from Kelud eruption at various locations can be summarized from a number of news online media reporting the eruption. Examples of reports that have been collected from the internet are as follows.

Reported by voaindonesia.com [22] at morning, Adisucipto airport in Yogyakartahas been covered by ash of 1-2 cm thickness. The visibility was less than 10 m. This ash thickness was deposited for about 4 hours.If the ash were fallen for 2 days 3 nigths as happened in Bima district during Tambora eruption, the ash thickness in Yogyakarta should become between 15 – 30 cm.

Reported by liputan6.com [23], at 6 AM (three hours after the ash fall), the ash rain continues to fell in Yogyakartaand has achive 3 cm thickness.Therefore, if we extrapolate to a period of 2 days and 3 nights, the ash thickness should become 60 cm. Reported in intisari-online.com, at morning of February 14, 2014 the ash thickness at Adi Sucipto airport, Yogyakarta was 2 cm at apron ann 6 cm at runway [24]. If we assume the condition was observed at around 9 AM, this thickness was deposited for 6 hours. If we extrapolate into 2 days and 3 nights, the thickness would become 20 cm – 60 cm. Other report by bpbd.slemankab.go.id, in Sleman district which is adjacent to Yogyakartawas was also covered by thick ash. Until 7.30AM, or around 4.5 hours after the ash started to fall, the ash thickness has achieved 5 cm [25]. If we extrapolate into 2 days and 3 nights the ash thickness would become 67 cm. In other report, at 9 AM, the thickness of ash in Adisucipto airport runway, Yogyakarta was around 6 cm [26]. Extrapolating into 2 days and 3 nights resulting a thickness of 60 cm.

However, the conditions of Yogyakarta and Sleman at the period of Kelud eruption were not as dark as Bima at the period of Tambora eruption. From television news it is apparent that the city conditions was still clear, at least at a short distance, and the resindents still able to see clearly within tens of meters. For example the following video in Youtube is the condition of Yogyakarta city at 7.28 AM on February 14, 2014 [27]. The Bima condition was very different, as express in Poem of Bima Kingdom, especially in strophes 23, 33, 36, and 41 of **Table 2**. Such strophes explained that the condition of Bima at day was nearly the same as night. Therefore, the thickness of ash in Bima nust be larger than in Yogyakarta and Sleman even after extrapolating into 2 days and 3 nights.

The conditions which nearly approached the Bima district during Tambora eruption was the condition of Kediri city, East Java, during Kelud eruption. This city is located 25 km from Kelud caldera. At morning, the city was likely dark. Based on TV news, the condition of Kediri city at morning was nearly the same as night [28]. Reported by solopos.com [29], Kediri city was covered by ash of 5 cm thickness after 3 hours hit by ash. Therefore, if we extrapolate into 2 days and 3 nights, the thickness of ash in Kediri city should become 100 cm. And the thickness of ash in Bima distcrict should be in the same order or larger. The most logical value to represent the ash thickness in Bima district was about the same as the thickness in Sumbawa or Lombok, i.e., around 60 cm. This opinion is based on the number of death in Bima and Sumbawa was nearly the same and the death in Lombok was smaller than in Bima.

Furthermore, the eruption of Kelud only happened at one day in the interval for a few hours. To the contrary, the mount Tambora eruption happened for almost one week. The eruption started at April 5, 1815 [8] and at the next morning ash started to fall in eastern Java. The sound was heard until Makassar (380 km), Batavia (1,260 km), and Ternate (1,400 km). In this first eruption, the ash certainly has fallen in Bima district. At around 7 PM of April 10, 1815, the eruption increased [6]. Huge eruptions occurred the night to April 11, 1815. The explosion sound was heard until Fort Marlborough, Bengkulen (1,800 km), Mukomuko (2,000 km), and Ternate (1,400 km). The smell of nitric was perceived until Batavia. Ashfall stopped in Java between 14-17 April 14-17, 1815 due to heavy rain and stoped in Sulawesi at April 15, 1815.

**IV. PROPOSAL OF ASYMMETRY GAUSSIAN FOR ASH DISTRIBUTION**

Experts have estimated the amount of ash produced by the Tambora eruption based on isopach map. The main assumption was that the ash distribution is asymmetry conical with the peak at the center of caldera. The cone base was taken to be the outer isopach curve. The volume of the cone is calculated using the formula $V = Ah/3$ wirh $A$ is the cone base area and $h$ is the cone height. Assuming asymmetry conical distribution, Stother obtained the cone volume of 135 km$^3$. By adding 10% to account for ashes scattered at a greater distance, the expected volume of ash produced by Tambora eruption was 150 km$^3$, or dense rock equivalet of 30 km$^3$ – 70 km$^3$ [8]. Similar approach was performed by Verbeek [11]. Verbeek obtained the cone volume of around 103 km$^3$. Then he assumed that ashes scattered at a greater distance caounted about 50% of the cone volume so that the esmiated ash volume produced by Tambora erupsion was 150 km$^3$. The value 150 km$^3$ is widely accepted at present [3,30,31,32].

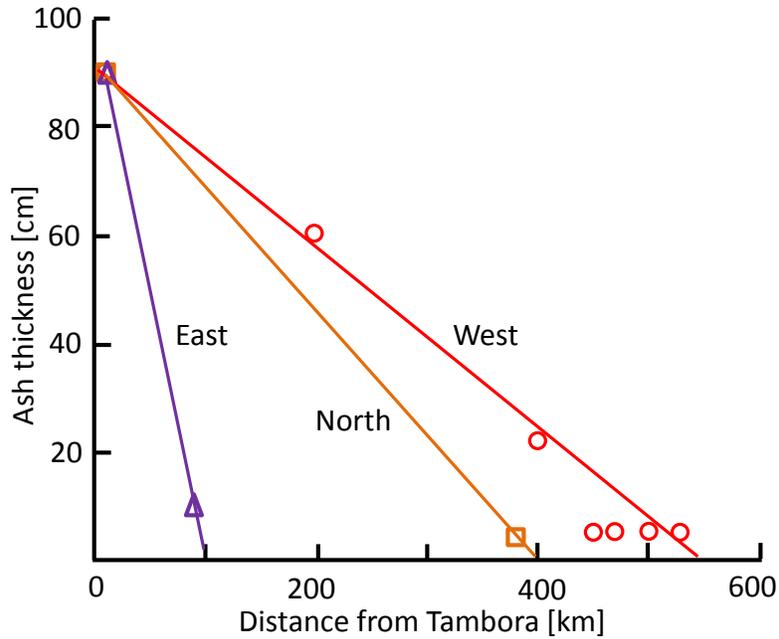

**Figure 3** Conical shroud lines that have been used by Strother to estimate the volume of ash produced by the Tambora eruption [8]

The next question was whether the asymmetry cone approximation correctly represents the distribution of Tambora ashes? **Figure 3** is the ash thickness data at various distances from Tambora toward west, north, and east directions. The lines represent cone shroud. If we pay attention to the west shroud line, it appears that the line does not accurately fit the thickness data. There are only one data at far distance toward the north and east so the straight line will always match the data. But we need to realize that generally natural processes will generally satisfy the normal distribution. Therefore, in this paper we assume that the distribution of Tambora ash satisfies the Gaussian distribution. But as the wind blew to the north-west, the ash was spread towards the north-west farther compared to the reverse direction so that we used an asymmetry Gaussian function to fit the ash ditribution much better. We used asymmetry Gaussian distribution as proposed by Kato et al [33]. Although there are several asymmetry Gaussian distributions, but we use the distribution function as proposed Kato et al because it is easier to calculate the volume. The thickness of the ash at coordinates x, and y is given by the equation

$$z(x,y) = z_0 \exp\left[-\frac{y^2}{2\sigma^2}\right] \times \begin{cases} \exp\left[-\frac{x^2}{2\sigma^2}\right] &, \quad x \geq 0 \\ \exp\left[-\frac{x^2}{2r^2\sigma^2}\right] &, \quad x < 0 \end{cases} \qquad (1)$$

with $z_0$ is the thickness at $x=y=0$ (center of caldera), $\sigma$ is the standard deviation of the symmetric normal distribution, $r$ is the skew paremeter, $x$ is taken to direct parallel to farthest distribution (west-north), $y$ directs perpendicular to x (inset in Figure 2), and $z$ is the ash thickness. The parameters $\sigma$ and r are determined by fitting the curve with data in isopach map. After the fitting is done, the volume of ash produced by Tambora eruption is given by

$$V = \int_{-\infty}^{\infty}\int_{-\infty}^{\infty} z(x,y)dxdx$$

$$= 2\pi z_0 \sigma^2 (1+r) \qquad (2)$$

We performed random simulations to support the assumption that the distribution of the ash thickness better match the Gaussian distribution and asymmetry Gaussian distribution if there is a wind. The method used is as follows. Suppose initially the particles are located above the center of caldera. We will simulate where the particle will touch the ground after a certain time interval. We modeled the ash particles has four possible directions of motion in the horizontal plane, i.e.: +x, -x, +y, dan –y. We assume that the probabilities of movements in four directions are all the same. To determine the direction of motion of a particle we generate a random numbers $p$ where $0 \leq p \leq 1$, then we do the following selection:

a) If $0 \leq p < 0.25$ then the particle moves toward $-x$

b) If $0.25 \leq p < 0.5$ then the particle moves toward $-y$

c) If $0.5 \leq p < 0.75$ then the particle moves toward $+y$

d) If $0.75 \leq p < 1$ then the particle moves toward $+x$

We only simulate particle motion in the x direction so the conditions that we use are only (a) and (d). Particles initially located at $x = 0$. Then we find the position x after a certain time interval.

We assume there is a wind blowing toward negative $x$ direction. Thus, the length of displacement toward the negative x direction is larger than that of toward the positive x direction. Thus, if $0 \leq p < 0.25$ the particle experiences a displacement of $-(1+a)$ with $0 \leq a \leq 1$. Conversely, if $0.75 \leq p < 1$ the particle experiences a displacement of $+(1-a)$. Such displacements are illustrated in **Figure 4**. We do simulation using Visual Basic.

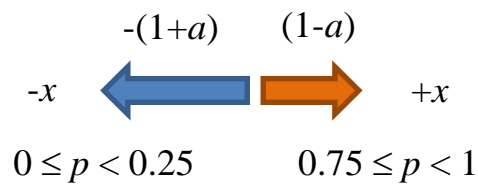

**Figure 4** Ilustration of ash particles displacements.

## V. CALCULATIONS

First we did simulations to show that the ash distribution better satisfies the Gaussian distribution rather than the cone distribution. We did simulations using 5000 particles and time duration of 4000 steps, either for no wind or with wind blowing toward the negative x direction. Figure **5** is the curve of ash thickness along the axis *x*. Figure **5(a)** is the curve without wind ($a = 0$) and Figure **5(b)** is the curve with wind ($a = 0.01$). It is clear that the ash thickness distribution better satisfies the Gaussian or asymmetry Gaussian distribution.

After proving that the distribution of ashes satisfies the asymmetry Gaussian distribution, then we did fitting to find parameters σ and *r*. For simplicity we did fitting along the x axis (y = 0). On this axis, equation (1) can be written as

$$z(x,0) = z_0 \begin{cases} \exp\left[-\dfrac{x^2}{2\sigma^2}\right] , & x \geq 0 \\ \exp\left[-\dfrac{x^2}{2r^2\sigma^2}\right] , & x < 0 \end{cases} \quad (3)$$

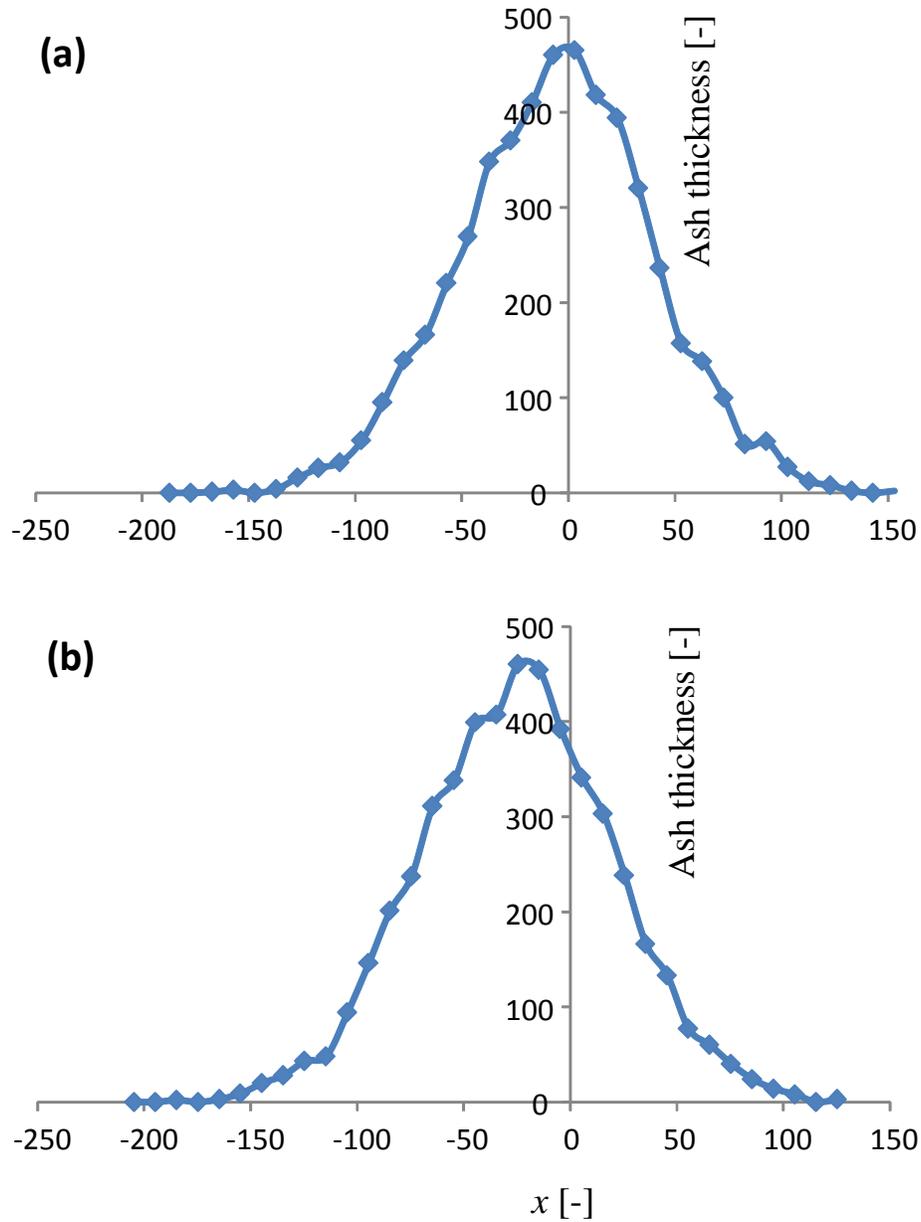

**Figure 5** The simulation results of Tambora ash thickness distribution along the x axis. The simulation was carried to account 5000 particles for 4000 steps. The horizontal and vertical axes on the graph are expressed in arbitrary units. We just want to show that the distribution of ash thickness better match the Gaussian distribution rather than cone distribution. Figure (a) for condition without wind and (b) for condition with wind blowing toward the negative x direction.

In the fitting process we use the thickness of ash in Bima district of 60 cm, equal to the thickness of the ash in Sumbawa and Lombok and is roughly equal to Francis note of 2 feet [17]. Distances of locations in isopach map (Figure 2) which coincides with the x-axis to the center of the caldera was determined using google map. **Table 4** is the positions along the x axis and the corresponding ash thicknesses.

**Table 4** The thickness of the ash at different coordinates extracted from Figure 2. The thickness of the ash in Bima district was taken equal to the thickness in Sumbawa or Lombok, i.e. 60 cm.

| Positions (km) | Location | Ash tihickness (cm) |
|---|---|---|
| 90 | Bima | 60 |
| 0 | Kaldera | 100 |
| -235 | North of Lombok island | 50 |
| -322 | North of Bali island | 25 |
| -531 | North of Madura island | 10 |
| -814 | Souht of Kalimantan island | 5 |
| -1092 | South of West Kalimantan | 1 |

**Figure 6** is the fitting result of ash thickness distribution along the x using equation (3). The best fitting was obtained using parameters $\sigma = 88$ km and $r = 2.4$. Using the obtained fitting parameters, we then predicted the ash volume produced during Tambora eruption, calculated using equation (2), as much as 165 km$^3$. If we add 10% to account for the ashes distributed at larger distances, we get the total volume of Tambora ash as 180 km$^3$. This figure is about 20% higher than the currently accepted value of 150 km$^3$.

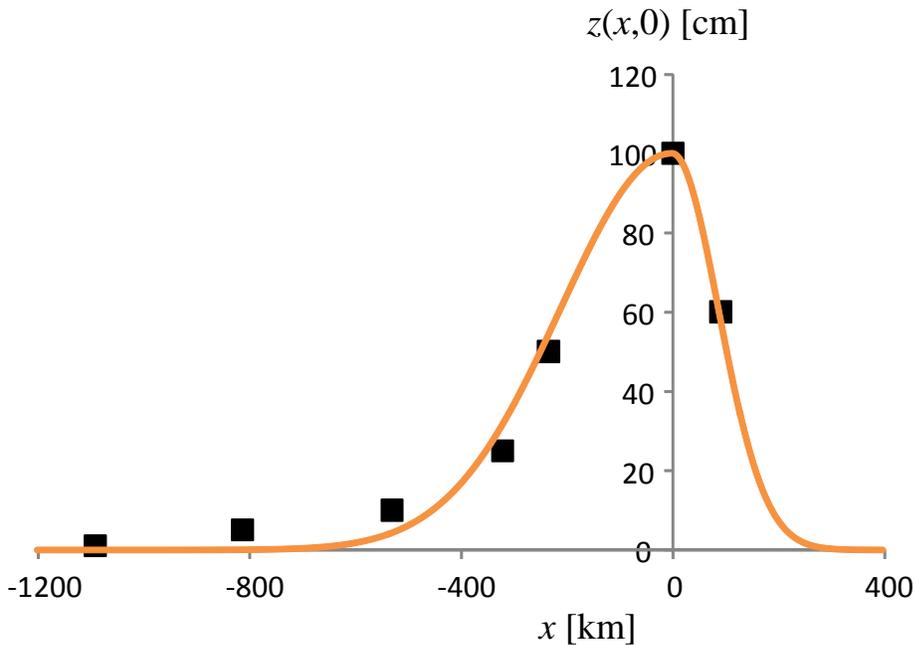

**Figure 6** The curve obtained from fitting the thickness ash distribution. The best fitting was obtained using $\sigma = 88$ km and $r = 2.4$ and resulted the ash volume of about 165 km$^3$.

**Figure 7** (top) is the profile of ash distribution calculated using Gaussian function (a) and asymmetry Gaussian function (b). Asymmetry Gaussian function is obtained from Eq. (1) by choosing $r = 1$. The (b) shape was assumed to represent the ash distribution from Tambora eruption. Figure **7** (bottom) is the profile of ash distribution calculated using symmetry cone (c) and asymmetry cone (d). It is seen a strong difference of ash distribution proposed here (b) and the distribution used by Strother (d).

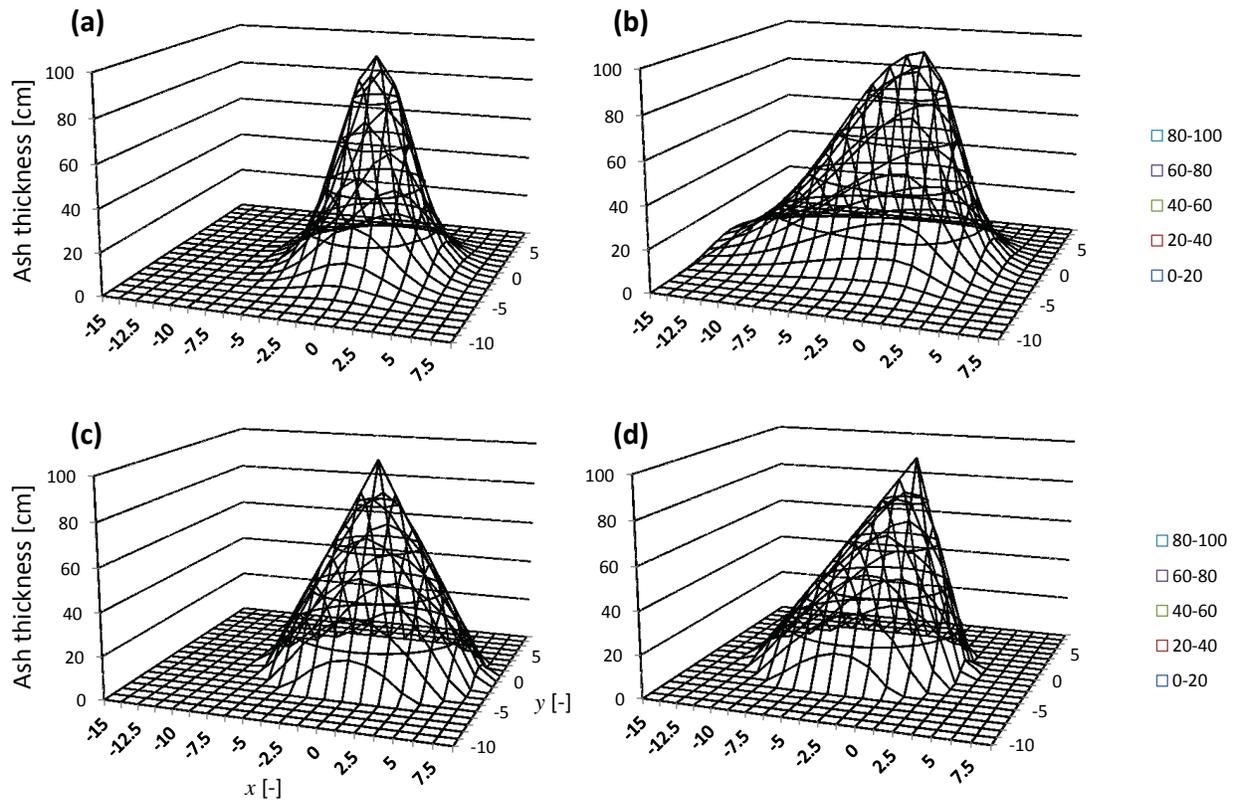

**Gambar 7** (a) Illustration of a symmetry normal distribution (b) an asymmetry normal distribution given by equation (1), (c) symmetry cone distribution, and (d) asymmetry cone distribution. Numbers on the axis are still in arbitrary unit. Figure (b) is the assumption used in this paper and Figure (d) is the distribution assumptions used by Strother [8].

**CONCLUSION**

I have reconstructed the distribution of ash released from great Tambora eruption in 1815. My reconstruction was developed by analyzing the meaning of *Poem of Bima Kingdom* written in 1830 and by comparing the effect of Tambora eruption and the Kelud eruption February 13, 2014. I get a more logical value for ash thickness in Bima district of around 60 cm, not 10 cm as widely reported. I proposed the ash distributed according to asymmetry Gaussian function. I have also done a random simulation to prove the existence of asymmetry Gaussian distribution. Simulation and direct fitting of the data extracted from isopach map showed that the asymmetry Gaussian distribution is more acceptable. I also obtained the new value for the total volume of ash released in Tambora eruption, i.e. around 180 km$^3$, greated of about 20% than the present accepted value of 150 km$^3$.

**REFERENCES**

[1] W. F. Libby, Atmospheric helium three and radiocarbon from cosmic radiation, Phys. Rev.69, 671–672 (1946).

[2] E. C. Anderson, W.F. Libby, S. Weinhouse, A.F. Reid, A.D. Kirshenbaum, and A.V. Grosse, Radiocarbon from cosmic radiation, Science 105, 576-577 (1947)

[3] P. Francis, Volcanoes: A Planetary Perspective (Oxford: Clarendon Press, 1993), p. 226.

[4] http://phys.org/news/2013-10-team-samalas-volcano-source-great.html

[5] en.wikipedia.org, 2016

[6] B. de Jong Boers, Mount Tambora In 1815: A Volcanic Eruption in Indonesia and Its Aftermath, Indonesia 60, 37-59 (1995)

[7] GoogleMap (2016)

[8] R. B. Stother, The Great Tambora Eruption in 1815 and Its Aftermath, Science 224, 1191-1198 (1984)

[9] F. Junghuhn, Java seine Gestalt, Pfanzendecke und innere Bauart, translated into Germany by J. K. Hasskarl (Arnold, Leipzig, 1854) vol. 2, p.818

[10] H. Zollinger, Besteigung des Vulkanes Tambora auf der Insel Sumbawa und Schilderung der Erupzion desselben im Jahr 1815 (Wurster, Winterthur, 1855)

[11] R.D.M. Verbeek, Nature (London) 30, 10 (1884)

[12] W.A. Petroeschevsky, Tijdschr. K. Ned. Aardrijks.Genoot. 66, 688 (1949)

[13] J.J. Pannekoek van Rheden, Z. Vulkanol. 4, 85 (1917)

[14] R.W. van Bemmelen, The Geology of Indonesia (Government Orinting Office, The Hague, 1949), vol. IA, p. 503

[15] Wikimedia.org

[16] See G. Kuperus, Het cultuurlandschap van West-Soembawa (Groningen: J. W. Wolter, 1936), pp. 22

[17] E. Francis, Herinneringen uit den levensloop van een Indisch ambtenaar van 1815 tot 1851 (Batavia, 1856), p. 137

[18] H. Chambert-Loir, Kerajaan Bima dalam Sastra dan Sejarah, Kepustakaan Populer Gramedia, Jakarta (2004), pp 219-378.

[19] C. G. C. Reinwardt, Reis naar het oostelijk gedeelte van den Indischen archipel, in het jaar 1821 (Amsterdam, 1858)

[20] W. van Bekkum, Geschiedenis van Manggarai (West-Flores), Cultureel Indie 8 (1946): 69

[21] M. Erb, When Rocks Were Young and Earth Was Soft: Ritual and Mythology in Northeastern Manggarai (PhD dissertation, State University of New York, 1982), p. 37

[22] http://www.voaindonesia.com/a/hujan-abu-gunung-kelud-sampai-yogyakarta-/1851206.html

[23] http://news.liputan6.com/read/826541/abu-gunung-kelud-sampai-yogyakarta-warga-rasakan-hawa-panas

[24] http://www.intisari-online.com/read/kelud-meletus-bandara-bandara-besar-tutup-sementara-

[25] http://bpbd.slemankab.go.id/?p=1758>

[26] http://adisutjipto-airport.co.id/detail/berita/akibat-letusan-gunung-kelud-bandara-adsutjipto-di-tutup-

[27] https://www.youtube.com/watch?v=4YtEXoHsKm8

[28] https://www.youtube.com/watch?v=pcLMDbX0OVA

[29] http://www.solopos.com/2014/02/14/gunung-kelud-meletus-kediri-diselimuti-abu-vulkanik-3-jam-ketebalan-5-cm-489409

[30] See W. A. Petroeschevsky, A Contribution to the Knowledge of the Gunung Tambora (Sumbawa), Tijdschrift van het Koninklijk Nederlandsch Aardrijkskundig Genootschapf 2nd ser., vol. 66 (1949): 695;

[31] M. R. Rampino and S. Self, Historic Eruptions of Tambora (1815), Krakatau (1883) and Agung (1963), Their Stratospheric Aerosols and Climatic Impact, Quaternary Research 18, 2 (1982): 129;

[32] see G. Neeb, The Composition and Distribution of the Samples, The Snellius-Expedition 1929-1930 5, 3, 2 (Leiden 1943): 88-97

[33] T. Kato, S. Omachi, and H. Aso, Assymteric Gaussian and Its Application to Pattern Recognition, Structural, Syntactic, and Statistical Pattern Recognition, Vol. 2396 of the series Lecture Notes in Computer Science, pp. 405-413 (2002)].